\documentclass[runningheads]{llncs}
\usepackage{graphicx}
\usepackage{cite}
\usepackage[colorlinks=true,linkcolor=blue,citecolor=blue,urlcolor=blue]{hyperref}
\usepackage[all]{hypcap}
\begin{document}
\title{Local Morphological Measures Confirm that Folding within Small Partitions of the Human Cortex Follows Universal Scaling Law}
\titlerunning{Folding within Small Partitions of the Cortex Follows Universal Scaling Law}
%
\author{Karoline Leiberg\inst{1} \and Christoforos Papasavvas\inst{1} \and
Yujiang Wang\inst{1}}
\authorrunning{K. Leiberg \textit{et al.}}
%
\institute{Newcastle University, Newcastle upon Tyne NE1 7RU, United Kingdom}
\maketitle              
\begin{abstract}
The universal scaling law of cortical morphology describes cortical folding as the covariance of average grey matter thickness, pial surface area, and exposed surface area. It applies for mammalian species, humans, and across lobes, however it remains to be shown that local cortical folding obeys the same rules. Here, we develop a method to obtain morphological measures for small regions across the cortex and correct surface areas by curvature to account for differences in patch size, resulting in a map of local morphology. It enables a near-pointwise analysis of morphological variables and their regional changes due to processes such as healthy ageing. We confirm empirically that the theorised covariance of morphological measures still holds at this level of local partition sizes as predicted, justifying the use of independent variables derived from the scaling law to identify regional differences in folding, subject-specific abnormalities, and local effects of ageing.

\end{abstract}
\section{Introduction}

Cortical morphology is a useful imaging-based biomarker for a range of applications, including aging and disease. Measures that describe the shape and folding of a brain, such as gyrification and cortical thickness, are being used both over the entire cortex and locally to identify differences in individuals and cohorts of subjects~\cite{Chaudhary2020,Frangou2021,Galovic2019,Libero2019}. Such measures are often studied separately, without taking their interaction into account.

Only recently, a universal scaling law describing the interaction of cortical morphology measures has been proposed~\cite{Mota2015}. This scaling law captures the folding of the cortex as the covariance between the measure of average cortical thickness $T$, total pial surface area $A_t$ and exposed surface area $A_e$ in the equation
\begin{equation}\label{ScalingLaw}
    A_t \sqrt{T} = k A_e^{1.25},
\end{equation}
where $k$ is a constant. Empirically, $k$ varies slightly between age groups, and has been interpreted as the tension/pressure applied to the cortical surface ~\cite{Wang2016}. This relation has been shown to hold for different mammalian species~\cite{Mota2015}, in human hemispheres~\cite{Wang2016}, and within human cortices across the lobes~\cite{Wang2019}. It says that the three variables are interdependent. For example, if two brains of a similar age have the same total area, but one is thicker, it will also have more exposed area - and hence less gyrification. 
The scaling law can be visualised by the analogy of crumpling a piece of paper into a ball: the thickness of the paper, its total area and the force used to crush it determine the exposed surface area of the ball~\cite{Mota2015}.

The scaling law additionally allows us to derive three linearly independent, interpretable morphological variables as linear combinations of $\log T^2$, $\log A_t$ and $\log A_e$~\cite{Wang2021}:
\begin{equation}\label{K}
    K = \log A_t + \frac{1}{4} \log T^2 - \frac{5}{4} \log A_e,
\end{equation}
\begin{equation}\label{I}
    I = \log A_t + \log T^2 + \log A_e,
\end{equation}
\begin{equation}\label{S}
    S = \frac{3}{2} \log A_t - \frac{9}{4} \log T^2 + \frac{3}{4} \log A_e.
\end{equation}
$K$ is derived directly from Eq.~(\ref{ScalingLaw}) by taking logarithms and arranging for $\log k$. It is a measure of tension/pressure acting on the cortex, or, in the analogy, pressure applied to the paper ball. $I$ describes the overall size of the brain or paper ball. Changes in $I$ correspond to an isometric scaling of the cortex. Lastly, $S$ is the inner product of $K$ and $I$ that contains information about shape; it can be thought of as the ``folding technique'' of the paper ball. Using $K$, $I$, and $S$ avoids covarying variables, whilst keeping them interpretable. Otherwise, analysing cortical thinning without accounting for surface area changes could miss signs of atrophy that can be discovered when accounting for the covariance~\cite{Wang2021}.

However, one distinct challenge remains in regionalising these independent measures of cortical morphology, $K$, $I$, and $S$. It is clear that the scaling law cannot hold for arbitrarily small regions. The goal of this paper is to show that the scaling law still holds for areas smaller than lobes, defined independently of them. We first develop a method to find the raw variables $T$, $A_t$, and $A_e$ in small patches across the cortex, and correct the surface areas so they are independent of the size chosen for patches. This allows us to compare the folding within the regions to that of other partitions of the cortex or even the full hemisphere~\cite{Wang2019}.

If local cortical folding follows the scaling law, it justifies the use of $K$, $I$, and $S$ for further morphological analyses. For example, we will show that they can be used to get a better understanding of regional age-related cortical atrophy.

\section{Methods}

\subsection{MRI data and processing}

To asses the rules of local cortical folding in a cohort of healthy subjects in a large age range, we used the NKI Rockland Sample data  (\url{http://fcon_1000.projects.nitrc.org/indi/pro/nki.html} \cite{Nooner2012}). 

The MRIs were preprocessed with the FreeSurfer 6.0 recon-all pipeline to obtain the fine, triangular mesh representing the grey matter surface, as well as the grey matter thickness on each point of the mesh. We then ran the local gyrification index processing stream to obtain the outer smooth (exposed) pial surface.

Out of the 929 subjects for whom data was available, 95 subjects were rejected due to inadequate image quality, motion artifacts or parts of their grey matter missing. We performed visual quality inspections on a random sample of 50 subjects and applied manual corrections where needed. We deemed overall image quality and FreeSurfer processing to be adequate. The remaining sample consists of 834 subjects, 508 female and 326 male, in an age range from 6 to 85 years. A full list of subjects used can be found on github (\url{https://github.com/KarolineLeiberg/folding_pointwise}).

\subsection{Extraction of local morphological measures}
The following describes the method in which morphological measures were computed for each point on the pial surface. The full code is available on github: \url{https://github.com/KarolineLeiberg/folding_pointwise}.

The pial surface and smooth pial surface are reduced to 5\% and 10\% of their original resolution respectively, using the downsampling function of the MATLAB iso2mesh toolbox. This retains the main features of cortical folds, but reduces the number of vertices in the pial to about 7000 per hemisphere.

The thickness map is converted from the original pial surface to the downsampled pial by assigning the value of the nearest pial vertex for each downsampled pial point. The pial thickness is subsequently transformed from being a pointwise measure to a measure for each face in the mesh (\url{https://github.com/cnnp-lab/CorticalFoldingAnalysisTools}).

\begin{figure}[htp]
\centering
\includegraphics[scale=0.55]{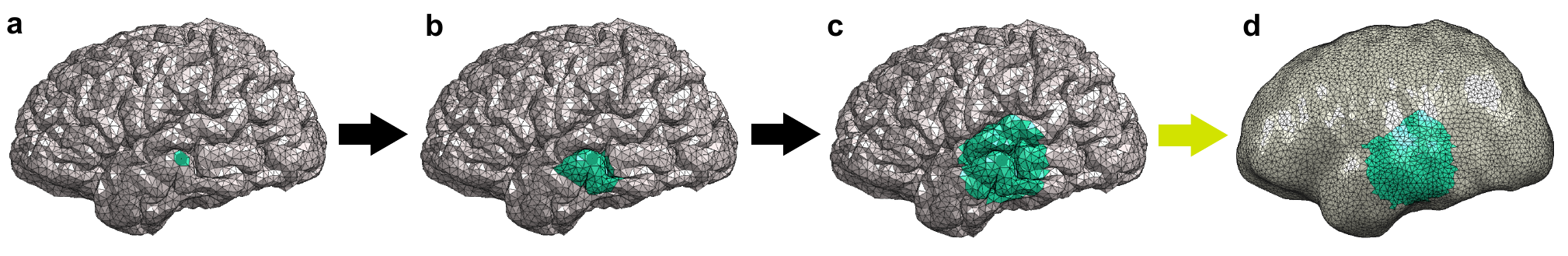}
\caption{Defining a patch on the pial around a point. \textbf{a-c} Extracts of the process of adding neighbouring points to the patch. \textbf{c} The final pial surface patch around this point. \textbf{d} The corresponding patch on the smooth pial.} \label{fig1}
\end{figure}

Then, for each point $p$ on the downsampled pial surface, a local patch around $p$ is defined as all vertices within a radius of 25mm, measured as Euclidean distance, that are directly connected to $p$ through neighbouring points which are also in the patch - this is to avoid e.g. disconnected patches across two gyri that are very close to each other, but the connecting sulcus is not included. A visualisation of the process can be found in Figure ~\ref{fig1} a-c. A corresponding smooth patch is defined by first finding the nearest downsampled pial point for each downsampled smooth pial point, and  then including smooth pial points if their neighbouring pial point is in the patch. (Fig.~\ref{fig1} d).

The total surface area $A_t^p$ for the patch around $p$ is then computed as the sum of all triangular faces of the downsampled pial which are contained in $p$'s patch. Similarly, the average cortical thickness $T^p$ is the average thickness across those faces. The exposed surface area $A_e^p$ is computed as the sum of face areas within the patch on the downsampled smooth pial.

\sloppy A convex hull is fitted over $p$'s downsampled smooth pial patch, and the integrated Gaussian Curvature of the patch around $p$ is approximated as the sum of Gaussian curvatures of all points on the convex hull which do not lie on the edge of the smooth patch (\url{https://github.com/cnnp-lab/CorticalFoldingAnalysisTools}).

Having computed the morphological measures for regions around each point on the downsampled pial surface, the data is converted back to the original, full pial, again using values of the nearest neighbour. This is done so the data can later be projected to the FreeSurfer average subject to compare across subjects.

\subsection{Surface area reconstruction}

As shown in Wang \textit{et al.} \cite{Wang2019}, the size of regions into which a cortex is partitioned fully determines its surface areas and thus its location in the space of $A_t$, $A_e$ and $T$, making it impossible to compare patches to each other and infer whether they align on the plane predicted by the scaling law. Each patch's surface areas are thus reconstructed to what they would be if the patch was a full hemisphere of the same gyrification index, using the proportion of curvature it contains, approximated via its convex hull. This correction is done using the formulas~\cite{Wang2019}
\begin{equation}\label{Atdash}
    A_t^{\prime p} = A_t^p * \frac{4 \pi}{I_G^p},
\end{equation}
\begin{equation}\label{Aedash}
    A_e^{\prime p} = A_e^p * \frac{4 \pi}{I_G^p},
\end{equation}
where $A_t^p$ and $A_e^p$ are the total and exposed surface areas of patch $p$ before correction, and $A_t^{\prime p}$ and $A_e^{\prime p}$ are the values after correction using the integrated Gaussian curvature $I_G^p$ over $p$. This correction of a patch's surface areas by its integrated Gaussian curvature preserves its average cortical thickness, gyrification index, and the average Gaussian curvatures of its total and exposed areas~\cite{Wang2019}.

Points that lie on particularly flat parts of the cortex or have very small surface areas tend to have convex hulls with integrated curvatures close to zero. In such cases, the correction of the surface areas leads to an overcorrection, inflating the surface area to unreasonably large values. For simplicity, we will exclude these points from our analysis, choosing not to investigate if they also obey the scaling law at this stage. All points within 10mm to the left and right of the midline of the brain and all points with a curvature below 0.16 are excluded. We will discuss later on how future work may be able to investigate these more challenging parts of the cortex, where the Gaussian curvature of the convex hull is not a good representation of the proportion of the patch of interest.

\subsection{Fitting the scaling law within subjects}

We assess if local folding follows the scaling law by regressing within each subject over all patches in the two-dimensional projection of the scaling law, where
\begin{equation}\label{X}
    X=\log A_e^{\prime p},
\end{equation}
\begin{equation}\label{Y}
    Y=\log (A_t^{\prime p}\sqrt{T^p}).
\end{equation}
Here, $A_t^{\prime p}$ and $A_e^{\prime p}$ are the surface areas after correcting by curvature, $T^p$ is the observed thickness. The slope of the regression is a subject-specific estimation of the exponent of $A_e$ in Eq.~(\ref{ScalingLaw}). We verify if each subject's local folding follows the scaling law by seeing how close the slope is to 1.25.

\subsection{Local age effects}

To quantify the effect ageing has on morphology in the raw measures and in the independent variables, we used the FreeSurfer function mri\textunderscore surf2surf to convert all subjects' morphology maps (i.e. pial surfaces with morphological measures computed at each point) to the same surface space, where they can be compared and analysed as a group. We compute the independent variables $K$, $I$, and $S$ on each point of the pial surface. We then fit linear regression models pointwise across all subjects, including both sex and age as covariates. Points represented in fewer than 10 subjects of either sex were excluded to ensure the regression was representative of the population. We then use the coefficient of the ageing covariate at each point and for each variable as an indicator of the local effect of age-related atrophy on that measure.

\section{Results}

\subsection{Surface area reconstruction by Gaussian curvature breaks down in insula and on midline}

\begin{figure}[htp]
\centering
\includegraphics[scale=0.5]{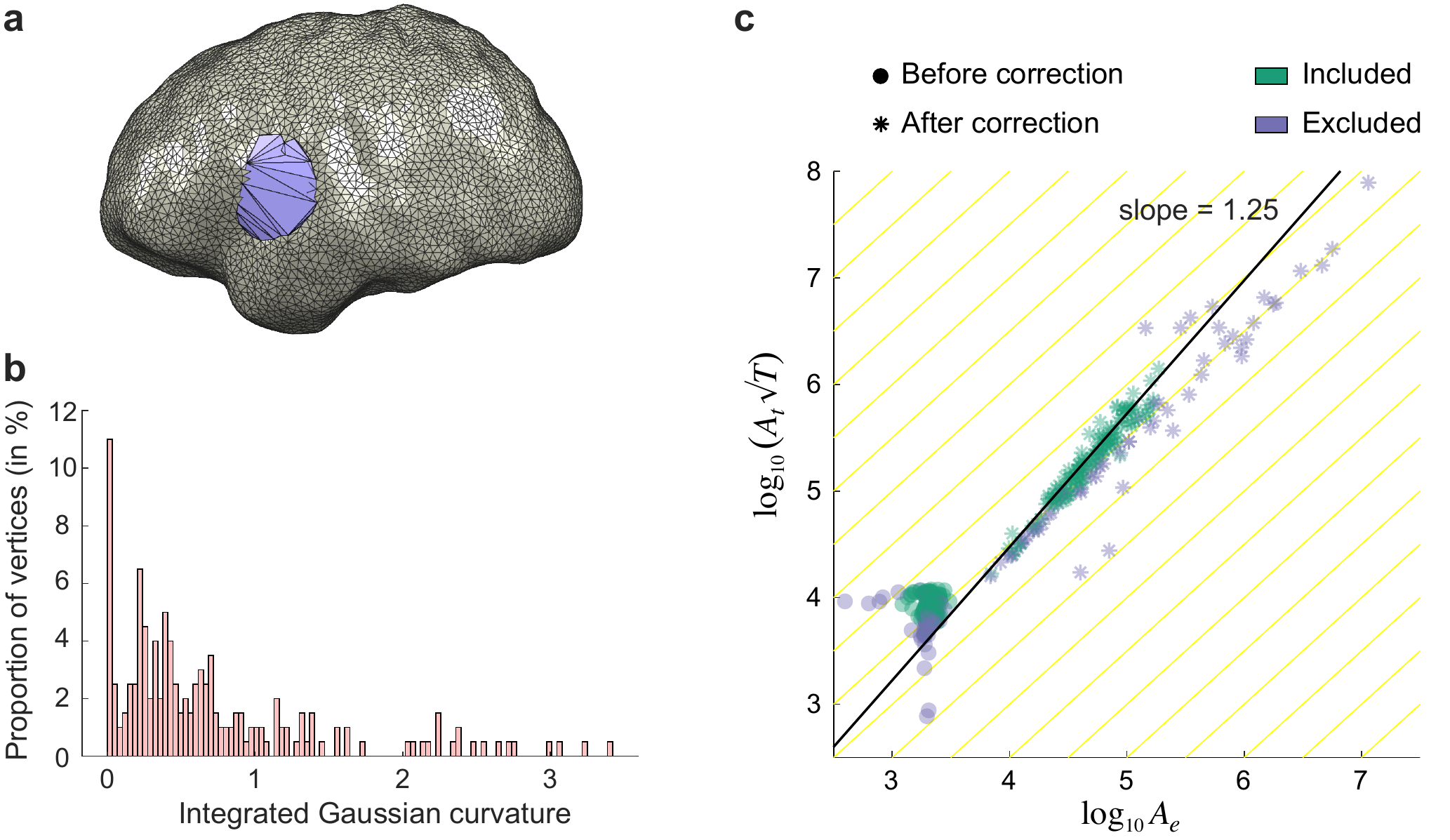}
\caption{Gaussian curvature and its effect on reconstructing surface areas in one example subject. \textbf{a} Example of patch with zero curvature. \textbf{b} Distribution of curvature across patches. \textbf{c} Sample of 150 patches plotted in the X-Y-plane, both before (round marker) and after (asterisk) surface area corrections. Yellow lines with slope 1 indicate the shift of points caused by the curvature corrections. Green points are included in further analysis, purple points are excluded due to their low curvature or position relative to the midline. The black line is a regression line through the green points after correction.} \label{fig2}
\end{figure}

We apply Gaussian curvature corrections (Eq.~(\ref{Atdash}),(\ref{Aedash})) to all points to reconstruct surface areas of patches to what they would be for a full hemisphere, but find that some points have approximately zero integrated Gaussian curvature (Fig.~\ref{fig2}a shows the convex hull over one example patch). When we look across all points of the cortex, we observe a distinct subset of points that display zero curvature (Fig.~\ref{fig2}b). These points are usually located in the insula or particularly deep sulci, where the patch has a small, overall convex, exposed surface, or on the midline, where the exposed surface might be large, but very flat. The midline additionally contains points with very large curvature, where the convex hull over a patch can contain an angle of 90° or more.

Correcting points by extremely small curvatures leads to overcorrections (Fig.~\ref{fig2}c): Whilst most points naturally align in the $X$-$Y$-plane (Eq.~(\ref{X}),(\ref{Y})) with a slope of 1.25 after the surface areas are reconstructed, points with curvatures near zero are being overcorrected to the top-right of the plot. For simplicity, we therefore exclude these points from our further analysis, acknowledging that Gaussian curvature of the convex hull is not a good representation of the proportion of those patches. This affects around 20\% of all points on the pial surface.

\subsection{Local folding follows scaling law}

\begin{figure}[htp]
\centering
\includegraphics[scale=0.6]{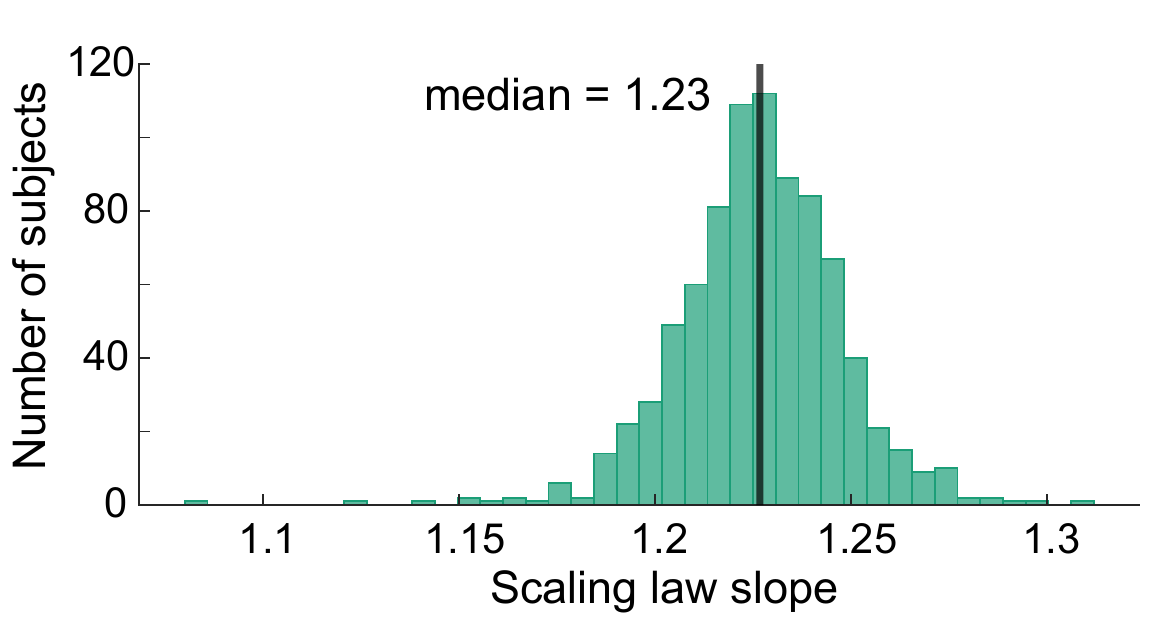}
\caption{Distribution of slopes observed in subjects when fitting a regression through points across the cortex in the $X$-$Y$-plane. The black line marks the median of observed slopes.} \label{fig3}
\end{figure}

To verify whether the scaling law still applies in small patches of the cortex, meaning if the measures of $T$, $A_t$, and $A_e$ covary locally as predicted, we fit the scaling law within subjects, estimating the slope between points in $X$ and $Y$ as described in Eq.~(\ref{X}) and (\ref{Y}). We find that the slope for each subject, i.e. the subject-specific estimates of the coefficient of $A_e$ in Eq.~(\ref{ScalingLaw}), are distributed around a mean of 1.23 (Fig.~\ref{fig3}), only slightly lower than the observed coefficient of 1.25 for hemispheres and lobes. This shows that the scaling law still holds at this level of patch sizes, meaning that when looking at small, local regions of the cortex with spherical radii of 25mm, the average thickness, total area and exposed area covary predictably according to Eq.~(\ref{ScalingLaw}). Note that the subjects follow the scaling law without any age or sex corrections, because age differences only affect the scaling law intercept ($K$), but not the slope~\cite{Wang2016}.

\subsection{$K$, $I$ and $S$ have additional value when observing ageing effects}

As we have seen, the folding follows the scaling law even locally, which implies that $K$, $I$, and $S$ are theoretically independent. We convert to these variables to see if they add insight to the differences in local morphology.

\begin{figure}[ht]
\centering
\includegraphics[scale=0.5]{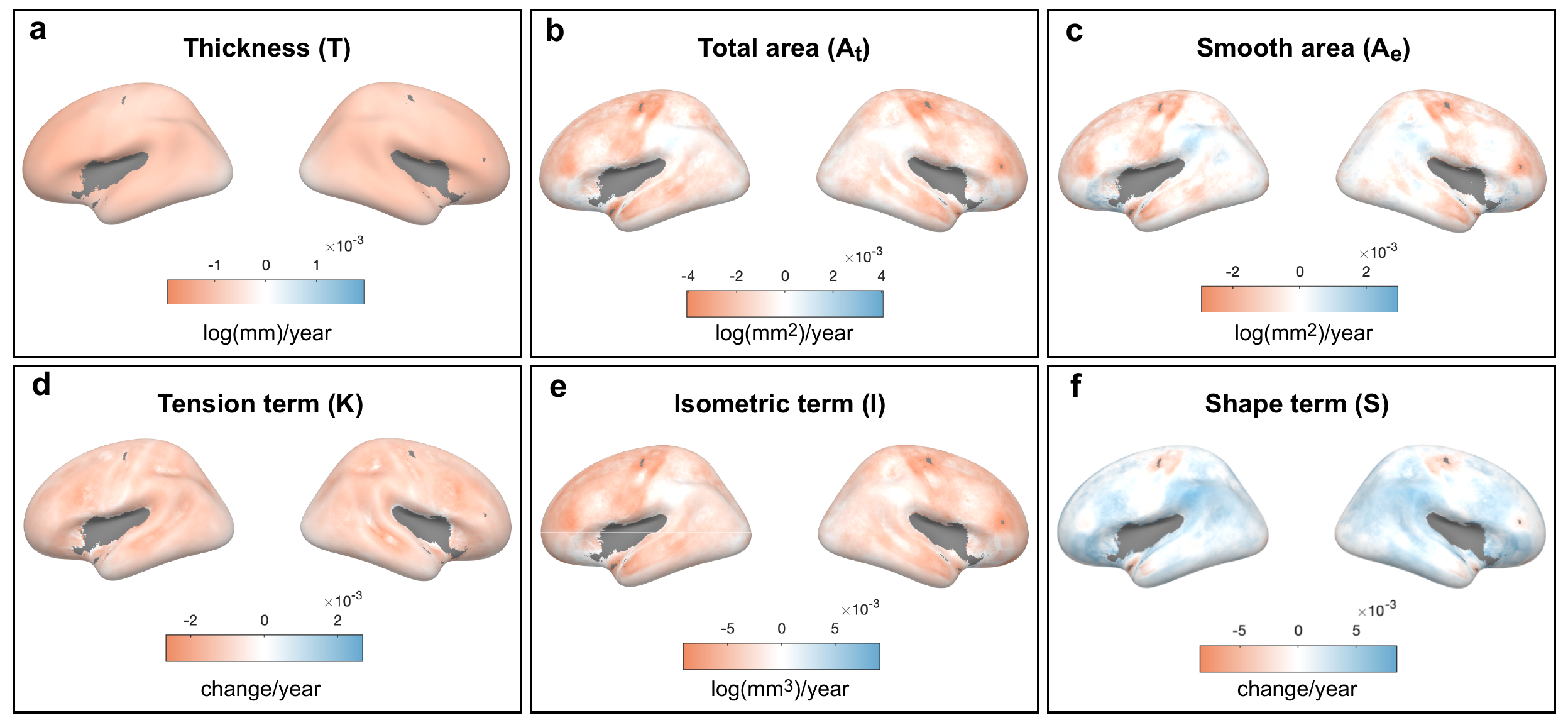}
\caption{Local effects of healthy ageing in the left and right hemisphere. \textbf{a-c} Raw variables $T$, $A_t$, and $A_e$ (logged). \textbf{d-f} Independent variables $K$, $I$, and $S$.} \label{fig4}
\end{figure}

When we look at the effect of healthy ageing on local morphology, we see a decrease in the raw variables $A_t$, $A_e$, and $T$ at varying rates in most areas of the brain (Fig.~\ref{fig4}~a-c). A decrease in surface areas can be interpreted as a flattening of the cortex, since less surface area within a constant radius indicates less folding. The independent variables show an isometric shrinking of most areas, with an overall loss of tension (Fig.~\ref{fig4}~d-e).

In the raw variables, we do not see systematic differences in how the areas around the upper and lower precentral gyrus are affected by atrophy. We do however see such a difference in the shape term $S$ (Fig.~\ref{fig4}f): $S$ decreases in the upper precentral gyrus with ageing, but the lower part is relatively unaffected. This is one example of the added value from switching to independent variables; we gain information otherwise hidden in the covariance of $A_t$, $A_e$, and $T$.

\section{Discussion}

On a big data set covering a large age range, we have demonstrated that the local folding of the brain follows the universal scaling law, meaning the grey matter thickness, total surface area and exposed surface area in small regions covary according to the same rule as the whole cortex. This result extends previous work done in this area, which had shown that morphology in full hemispheres and lobes adheres to the scaling law, by confirming empirically that the minimum size of regions for which the folding rule applies has a radius of under 25mm.

Our method works well for most of the cortex. However, we had to exclude around 20\% of points on the pial surface from our analysis, predominantly located on the midline and in the insula. For these points, we founds that the Gaussian curvature was not an adequate representation of the proportion of the patch. For future iterations of the method, we plan to improve on this. An alternative way to approximate patch curvature would be to compute the integrated curvature of the patch's smooth pial surface, rather than its convex hull. However, this surface may already contain too much local information of folding for our purpose. Another idea is to inflate the pial surface to a sphere and use the proportion of surface area or curvature of the patch's representation on the sphere for corrections.

Another improvement of the method could be achieved by making the radius defining the patch around each point on the pial adaptive depending on the thickness at that point. This flexible way of patch definitions uses the smallest radius required to find enough exposed surface area at each point, making the morphological map as close to pointwise as possible. We expect the ideal patch radii to be around 20-30mm for human brains (spanning at least one sulcus/gyrus), which is why we chose a value in this range for the fixed-size method.

The local applicability of the scaling law allows us to use it in terms of independent components derived from it, to further analyse local morphology. In our results we have shown local differences in how the cortex is affected by the process of ageing in the independent variables, which were not clear from the raw variables alone. Our method could have other practical applications, such as finding local abnormalities in patient groups compared to a control cohort, indicating regional effects of dysfunction. It might also be used to detect subject-specific abnormalities, indicating areas of the brain that fold atypically.

\subsubsection{Acknowledgements.} KL was supported by the Centre for Doctoral Training in Cloud Computing for Big Data (EP/L015358/1). We thank the members of CNNP lab (www.cnnp‐lab.com) and Bruno Mota for discussions of the method and results.

%
%
%
\bibliographystyle{splncs04}
\bibliography{Pointwise_folding}
%

\end{document}